\documentclass{nature}
\usepackage[pdftex]{graphicx}
\title{A Learning Approach to Optical Tomography}

\author{Morteza H. Shoreh$^{*,1}$, Ulugbek S. Kamilov$^{*,2}$, Ioannis N. Papadopoulos$^{*,1}$, 
	Alexandre Goy$^{1}$, Cedric Vonesch$^{2}$,
	Michael Unser$^{2}$ \& Demetri Psaltis$^{1}$}

\begin{document}

\maketitle
  \\ $^*$\textit{These authors contributed equally to the paper}
\begin{affiliations}

 \item Optics Laboratory, 	$\acute{E}$cole polytechnique f$\acute{e}$d$\acute{e}$rale de Lausanne, Switzerland  ́
  \item Biomedical Imaging Group, $\acute{E}$cole polytechnique f$\acute{e}$d$\acute{e}$rale de Lausanne, Switzerland  ́
\end{affiliations}

\begin{abstract}
We describe a method for imaging 3D objects in a tomographic configuration implemented by training an artificial neural network to reproduce the complex amplitude of the experimentally measured scattered light. The network is designed such that the voxel values of the refractive index of the 3D object are the variables that are adapted during the training process. We demonstrate the method experimentally by forming images of the 3D refractive index distribution of cells.
\end{abstract}

The learning approach to imaging we describe in this paper is related to adaptive techniques in phased antenna arrays\cite{Widrow} and inverse scattering\cite{scattering1, scattering2}. In the optical domain an iterative approach was demonstrated by the Sentenac group\cite{frechgroup1,frechgroup2} who used the coupled dipole approximation\cite{dipole} for modelling light propagation in inhomogeneous media (a very accurate method but computationally intensive) to simulate light scattering from small objects ($1\mu m\times 0.5 \mu m$) in a point scanning microscope configuration. Very recently an iterative optimization method was demonstrated\cite{Optica} for imaging 3D objects using incoherent illumination. Our method relies on digital holography\cite{holography1,holography2}  to record the complex amplitude of the field. We use the Beam Propagation Method (BPM)\cite{BPM,Goodman} to model the scattering process and the error back propagation method\cite{backpro} to train the system. At the end of the training process the network discovers a 3D index distribution that is consistent with the experimental observations. We experimentally demonstrate the technique by imaging polystyrene beads and HeLa and hTERT-RPE1 cells. 

\begin{figure*}[b!]
	\centering
	\includegraphics[width=5 in]{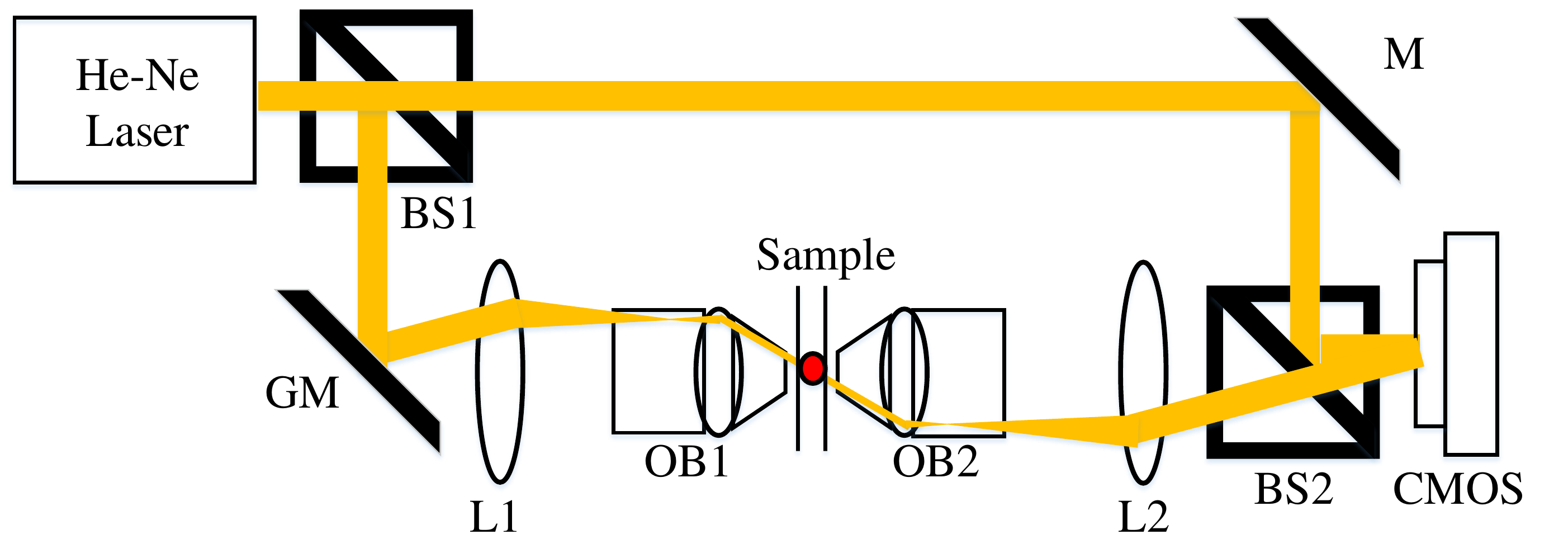}
	\caption{Experimental setup (BS: Beam Splitter, GM: Galva Mirror, L: Lens, OB: Objective, M: Mirror)}
\end{figure*}

\section*{Experimental Setup} 

A schematic diagram of the experimental setup is shown in Figure 1. It is a holographic tomography system\cite{Wolf}, in which the sample is illuminated with multiple angles and the scattered light is holographically recorded. Several variation of the holographic tomography system have been demonstrated before\cite{Choi1, Choi2, Choi3, setup}. The optical arrangement we used is most similar to the one described by Choi \textit{et al.}\cite{Choi1}. The samples to be measured were prepared by placing polystyrene beads and cells between two glass cover slides. The samples were illuminated with a continuous collimated wave at $561nm$ at 80 different angles. The amplitude and phase of the light transmitted through the sample was imaged onto a 2D detector where it was holographically recorded by introducing a reference beam. These recordings constitute the training set with which we train the computational model that simulates the experimental setup. We construct the network using the BPM. The inhomogeneous medium (beads or cells) is divided into thin slices along the propagation direction ($z$).  The propagation through each slice is calculated as a phase modulation due to the local transverse index variation followed by propagation in a thin slice of a homogenous medium having the average value of the index of refraction of the sample.

\section*{Methodology}
A schematic description of the BPM simulation is shown in Figure 2.  The straight lines connecting any two circles represent multiplication of the output of the unit located in the $l$-th layer of the network at $x=n_1\delta, y=m_1\delta$ by the discretized Fresnel diffraction kernel $e^{j\pi [(n_l^2-n_{l+1}^2)\delta^2+(m_l^2-m_{l+1}^2)\delta^2] /\lambda \delta_z}$  where $n_l$ and $m_l$ are integers and $\lambda$ is the wavelength of light. $\delta$ is the sampling interval in the transverse coordinates $(x ,y)$ whereas $\delta_z$ is the sampling interval along the propagation direction $z$. The circles in the diagram of Figure 2 perform a summation of the complex amplitude of the signals converging to each circle and also multiplication of this sum by $e^{j(2\pi\Delta n \delta_z z) /\lambda}$. $\Delta n(x,y,z)$  is the unknown 3D index perturbation of the object. 
\begin{figure*}[t!]
	\centering
	\includegraphics[width=6 in]{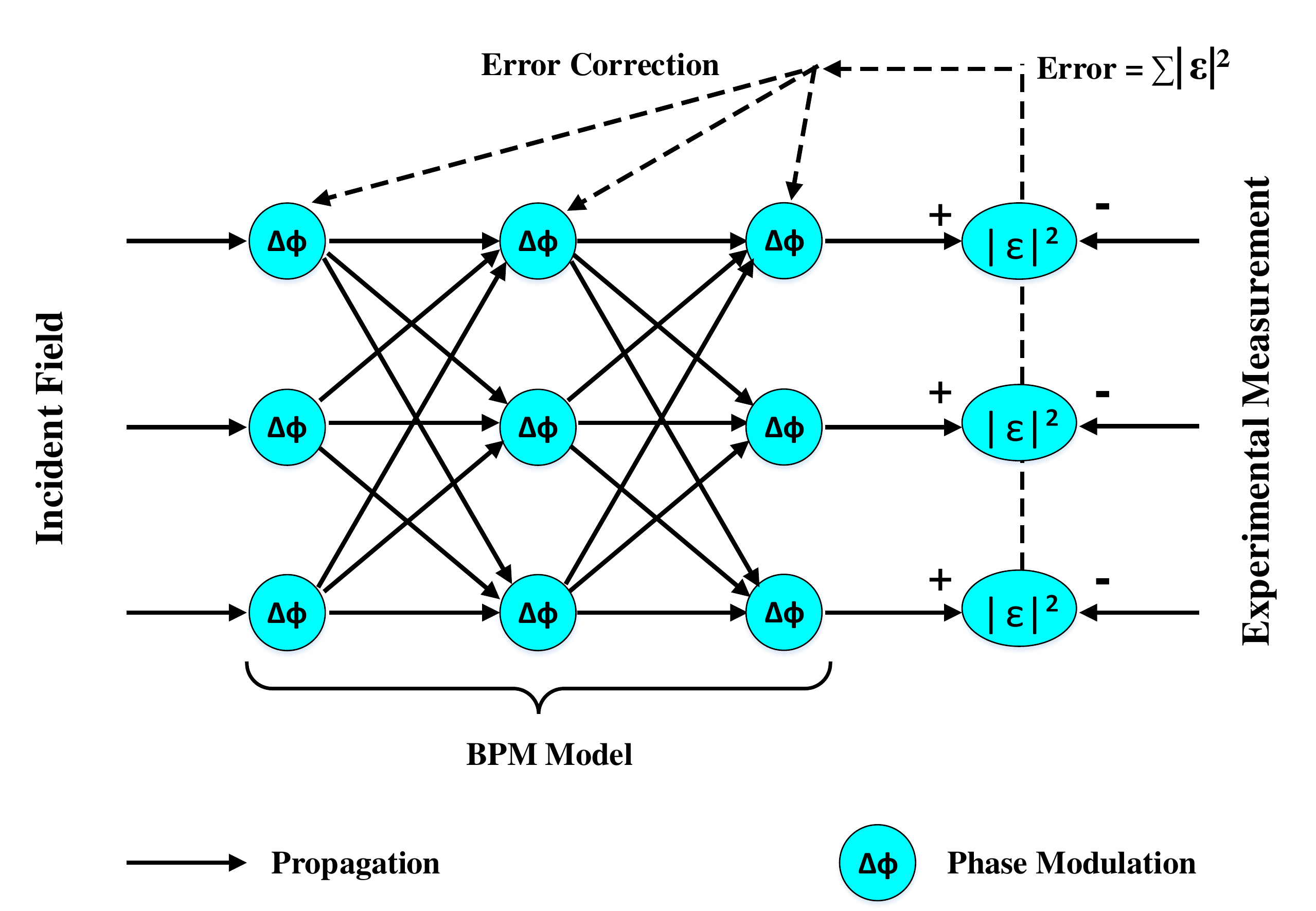}
	\caption{Schematic diagram of object reconstruction by learning the 3D index distribution that minimizes the error $\epsilon$, defined at the mean squared difference between the experimental measurement and the prediction of a computational model based on the beam propagation method (BPM).}
	\hspace{0.3 in}
\end{figure*}

In the experiments the network has 420 layers with $\Delta n(x,y,z)$ being the adaptable variable. In contrast with a conventional neural network, the output of the layered structure in Figure 2 is a linear function of the input complex field amplitude. However, the dependence of the output is nonlinearly related to $\Delta n(x,y,z)$. The BPM can be trained using steepest descent exactly as the back propagation algorithm in neural networks\cite{Gradient,Neural1, Neural2}. Specifically, the learning algorithm carries out the following minimization:
$$ {\min_{\Delta \hat{n}}}\{\frac{1}{2k} \sum_{k=1}^K \| E_k (\Delta \hat{n}) - M_k (\Delta n )\|^2 +\tau S(\Delta \hat{n})\}$$ $$
 \textit{subject to}~~0 \leq \Delta\hat{n}$$ 
In the above expression $E_k (\Delta \hat{n})$ is the current prediction of the BPM network for the output when the system is illuminated with the $k$-th beam and $M_k (\Delta n )$ is the actual measurement obtained by the optical system. $\Delta \hat{n}$ indicates the estimate for the index perturbation due to the object. The term $S(\Delta \hat{n})$ is a sparsity constraint \cite{Ulegbek} to enhance the contrast while $\tau$ is a parameter that can be tuned to maximize image quality by systematic search. The positivity constraint takes advantage of the assumption that the index perturbation is real and positive. The optimization is carried out iteratively by taking the derivative of the error with respect to each of the adaptable parameters following steepest descent
$$ \Delta \hat{n} \rightarrow\Delta \hat{n}-\frac{\alpha}{k} \sum_{k=1}^K \epsilon_k \frac{\partial \epsilon_k}{\partial\Delta\hat{n}} +\tau \frac{\partial S(\Delta \hat{n})}{\partial\Delta\hat{n}}$$ 
where $\epsilon_k=\| E_k (\Delta \hat{n}) - M_k (\Delta n )\|$ is the error, $\alpha$ is a constant and the change in $\Delta\hat{n}$ is proportional to the error and its derivative. This is achieved efficiently via a recursive computation of the gradient, which is the back propagation part of our learning algorithm.

\section*{Results}
We first tested the system with polystyrene beads encapsulated between two glass slides in immersion oil. The sample was inserted in the optical system of Figure 1 and 80 holograms were recorded by illuminating the sample at 80 distinct angles uniformly distributed in the range -45 degrees to +45 degrees. The collected data is the training set for the 420-layer BPM network which simulates a physical propagation distance of $30\mu m$ and transverse window $37\mu m \times 37\mu m$ ($\delta_x = \delta_y = 72 nm$).  The network was initialized with the standard filtered back projection reconstruction algorithm (Radon transform)\cite{Radon} and the resulting 3D images before and after 100 iterations are shown in Figure 3. The final image produced by the learning algorithm is an accurate reproduction of the bead shape. 

\begin{figure*}
	\centering
	\includegraphics[width=4.5in]{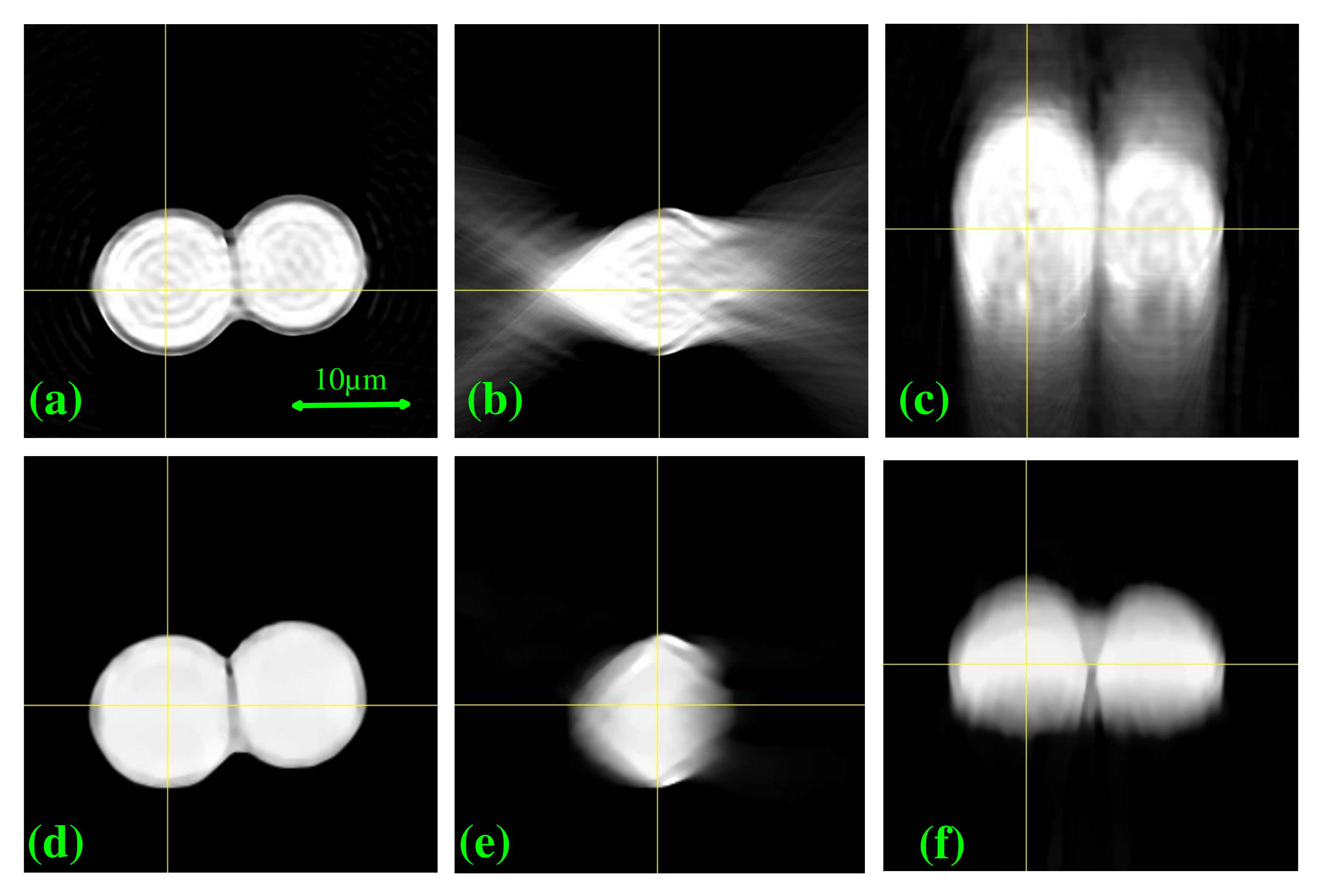}
	\caption{Reconstruction of two $10\mu m$ beads of refractive index $1.588$ at $\lambda=561nm$ in immersion oil with $n_0=1.516$. (a)-(c) x-y, y-z and x-z slices using the inverse Radon transform reconstruction, (d)-(f) the same slices for our learning based reconstruction method.}
	\hspace{0.3 in}
\end{figure*}

A sample of a HeLa cell was also prepared and the same procedure was followed to obtain a 3D image. The results are shown in Figure 4 where the error function is plotted as a function of iteration number.  In this instance, the system was initialized with a constant but nonzero value ($\Delta \hat{n}=0.007$). Also shown in Figure 4 are the results obtained when the system was initialized with the Radon reconstruction from the same data. After 100 iterations both runs yield essentially identical results. Notice that the error in the final image (after 100 iterations) is significantly lower than to the error of the Radon reconstruction. This is also evident by visual inspection of the images in Figure 4 where the artifacts due to the missing cone\cite{Cone} and diffraction\cite{Choi1} are removed by the learning process. 

\begin{figure*}
	\centering
	\includegraphics[width=4.9 in]{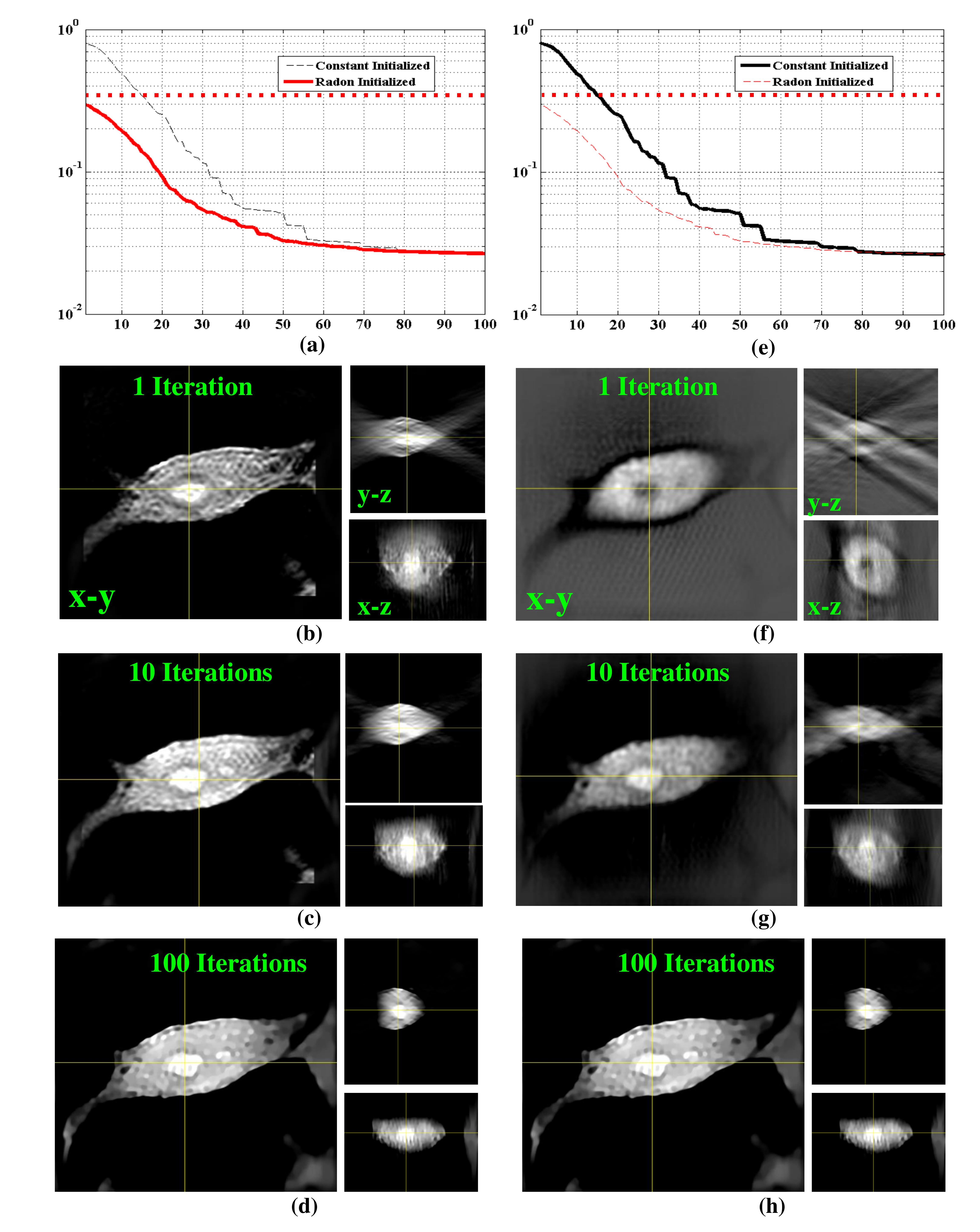}
	\caption{Comparison of the proposed method initialized by inverse Radon transform (left) versus initialization with a constant value ($\Delta\hat{n}=0.007$) (right), (a),(e) pixel error energy fall-off for initialized with inverse Radon and constant value, respectively. The horizontal doted line shows the inverse Radon performance for comparison. (b)-(d), x-y, y-z and x-z stacks for respectively the first, tenth and hundredth iteration of the proposed method initialized by inverse Radon. (d)-(f), the same figures for the proposed method initialized by constant value.}
	\hspace{0.3 in}
\end{figure*}


In general, optical 3D imaging techniques rely on the assumption that the object being imaged does not significantly distort the illuminating beam. This is assumed for example in Radon or diffraction/holographic tomography. In other words, these 3D reconstruction methods rely on the assumption that the measured scattered light consists of photons that have only been scattered once before they reach the detector. The BPM, on the other hand, allows for multiple forward scattering events. The only simplification is that reflections are not taken into account; these could eventually be incorporated in the network equation without fundamentally altering the approach described in this paper. Since biological tissue is generally forward scattering, BPM can be a good candidate to model propagation of thick biological samples and this may be the most significant advantage of the learning approach. To demonstrate this point, we prepared two glass slides with a random distribution of hTERT-RPE1 cells (immortalized epithelial cells from retina) on each slide. When we attach the two slides together, we can find locations where two cells are aligned in $z$, one on top of the other. Figure 5 (a)-(e) shows the image of such a stack of two cells produced with a direct inversion using the Radon transform. Figure 5 (f)-(j) shows the same object imaged with the proposed learning algorithm. The learning method was able to distinguish the two cells where the Radon reconstruction merged the two into a single pattern due to the blurring in $z$ which is a consequence of the missing cone. 

\begin{figure*}
	\centering
	\includegraphics[width=3.2 in]{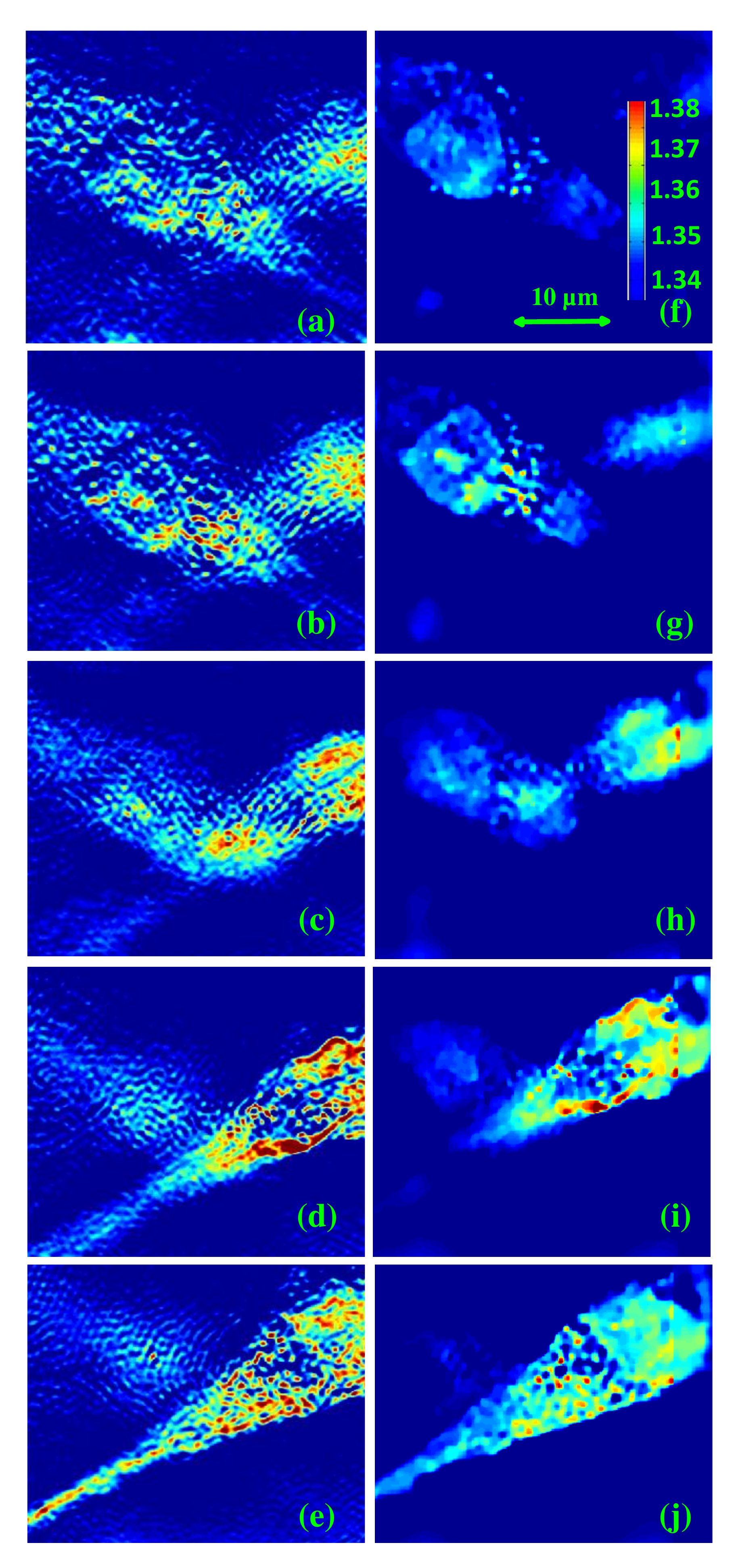}
	\caption{Images of two hTERT-RPE1 cells. x-y slices corresponding to different depths of respectively +9, +6, +3, 0 and -3 microns (positive being toward the detector) from the focal plane of the lens OB2 in Figure 1 for: (a)-(e) the inverse Radon transform based reconstruction and (f)-(j) the same slices for our learning based reconstruction method.}
	\hspace{0.3 in}
\end{figure*}

In conclusion, we have demonstrated a neural-network based algorithm to solve the optical phase tomography problem and have applied it to biological (HeLa and hTERT-RPE1 cells) and synthetic (polystyrene beads) samples. The experimental measurements were performed with a conventional collimated illumination phase tomography setup, with coherent light and holograms were recorded off-axis. The sample scattering potential was modeled as a neural network implementing a forward beam propagation method. The network is organised in neurones layers, each one of them representing an x-y plane in the BPM. The output of the network is compared to the experimental measurements and the error is used to correct the weights (representing the refractive index contrast) in the neurones using standard error back propagation techniques. The algorithm yields images of better quality than tomographic reconstruction (Radon). In particular, the missing cone artefact is efficiently removed, as well as parasitic granular structure. We have shown that whether starting from a constant initial guess for the refractive index or with a conventional Radon tomographic image, the method essentially converges to the same result after 100 iterations. This approach opens rich perspectives for active correction of scattering in biological sample; in particular, it has the potential of increasing the resolution and the contrast in fluorescent and two-photon imaging.




\begin{addendum}
 \item The authors would like to thank Phelps Edward Allen and Val$\acute{e}$rian CR Dormoy for sample preparation and Nicolino Stasio, Donald Conkey and Ye Pu for their helpful suggestions.
 \item[Competing Interests] The authors declare that they have no
competing financial interests.
 \item[Correspondence] Correspondence and requests for materials
should be addressed to Demetri Psaltis~(email: demetri.psaltis@epfl.ch).
\end{addendum}

\end{document}